\documentclass[12pt,a4paper,twoside]{article}
\def\runninghead#1#2{\pagestyle{myheadings}
\markboth{{\protect\footnotesize\it{\quad #1}}\hfill}
{\hfill{\protect\footnotesize\it{#2\quad}}}} \headsep=15pt
\usepackage{amssymb}
\usepackage{latexsym}
\usepackage{setspace}
\newcommand{\be}{\begin{equation}}
\newcommand{\ee}{\end{equation}}
\newcommand{\bea}{\begin{eqnarray}}
\newcommand{\eea}{\end{eqnarray}} %
 \title{Torsion, Dirac Field, Dark Matter and Dark Radiation}
 \author{Prasanta Mahato\footnote{Narasinha Dutt College,
         Howrah, West Bengal, India 711 101}\hspace{2mm}\footnote{E-mail: pmahato@dataone.in}}
\date{}
\begin{document}
\runninghead{\underline{Prasanta Mahato}}{\underline{Torsion, Dirac Field, Dark Matter and Dark Radiation}}
\maketitle
\setcounter{footnote}{0}
\singlespacing
 \begin{abstract}
 The role of  torsion and a scalar field $\phi$
   in gravitation, especially, in the presence of a Dirac field  in the background of a particular class of the Riemann-Cartan geometry is considered here.
 Recently,  a Lagrangian density with Lagrange multipliers has been proposed by the author which has been obtained  by picking some particular terms from the $SO(4,1)$ Pontryagin  density,  where the scalar field $\phi$ causes  the de Sitter connection to have the proper dimension of a gauge field. In this article the scalar field has been linked  to the dimension of the Dirac field. Here   we get the field equations for the Dirac field and  the scalar field  in such a way that both of them appear to be mutually non-interacting. In this scenario the scalar field appears to be a natural candidate for the dark matter and the dark radiation.

\vspace{2mm}
   \noindent \textbf{\uppercase{pacs:}} 04.20.Fy, 04.20.Cv
   
\vspace{2mm}
   \noindent \textbf{\uppercase{key words:}}   Nieh-Yan Density, Torsion, Dark Matter, Dark Radiation

 \end{abstract}
\doublespacing
\section{Introduction}

Scalar-Tensor gravity theories\cite{Ber68,Wag70,Nor70} provide a
natural generalizations of General Relativity (GR) by introducing
an additional scalar field $\phi$, the dynamical importance of
which is determined by an arbitrary coupling function
$\omega(\phi)$. According to Brans-Dicke
  theory, the value of $G$ is determined by the value of the Brans-Dicke scalar field
  $\phi$. The Brans-Dicke version of Einstein-Cartan theory, with nonzero torsion
  and vanishing non-metricity, has been discussed by many  authors\cite{Rau84,Ger85,Kim86}.
   In these approaches $\phi$ acts as a source of torsion\cite{Ber93,Der01}. On the other hand
String theories\cite{Gre88}, also, predict the existence of scalar partners to the tensor gravity of GR.
  Since  the  advent   of  the  first   inflationary  models
\cite{Linde},  cosmologies  containing   scalar  fields  have
received  widespread  attention  in  the literature.   From  a  purely
phenomenological point of view,  such scalar fields are general enough
to accommodate a rich variety  of behaviors.  From a theoretical point
of  view, their invariable  appearance in  various theories  of nature
makes them natural candidates for cosmological applications. Such fields
 as correspondent potential
in Riemann-Cartan FRW universe has also been studied in some other work\cite{Odi93}.

  It is a remarkable result of differential geometry that
 certain global features of a manifold are determined by some local invariant densities.
These topological invariants have  an important property in common -  they are total divergences
and in any local theory  these invariants,  when treated as Lagrangian densities, contribute
 nothing to the Euler-Lagrange equations.  Hence in a local theory only few parts, not the
  whole part, of these invariants can be kept in a Lagrangian density. Recently, in
this direction, a gravitational Lagrangian has been
proposed\cite{Mah02a}, where a
 Lorentz invariant part of the de Sitter Pontryagin density has been treated as
 the Einstein-Hilbert Lagrangian.  By this way the role of torsion in the underlying manifold
has become multiplicative
   rather than additive one and  the  Lagrangian  looks like
    $torsion \otimes curvature$.  In other words - the additive torsion is decoupled from the
   theory but not the multiplicative one. This indicates that torsion is uniformly nonzero
   everywhere. In the geometrical sense, this implies that
   micro local space-time is such that at every point there is a
   direction vector (vortex line) attached to it. This effectively
   corresponds to the non commutative geometry having the manifold
   $M_{4}\times Z_{2}$, where the discrete space $Z_{2}$ is just
   not the two point space\cite{Con94} but appears as an attached direction vector. This has direct relevance in the quantization of a fermion where the discrete space appears as the internal space of a particle\cite{Gho00}. Considering torsion and torsion-less
  connection as independent fields\cite{Mah04}, it has been found that $\kappa$  of Einstein-Hilbert Lagrangian, appears as an integration constant in such a way that it has been found to be linked with the topological Nieh-Yan density of $U_{4}$ space.
  If we consider axial vector torsion together with a scalar field $\phi$ connected to a local  scale factor\cite{Mah05}, then the Euler-Lagrange equations not only give the constancy of the gravitational constant but they also link, in laboratory scale, the mass of the scalar field with the Nieh-Yan density and, in cosmic scale of FRW-cosmology, they predict only three kinds of the phenomenological energy density representing mass, radiation and cosmological constant.

  This article has been prepared to study the role of  torsion and the scalar field $\phi$
   in gravitation, especially, in the presence of a Dirac field  in the background of  a particular class of Riemann-Cartan geometry of the space-time manifold and also try to study the possible connection of the field $\phi$ with dark matter\cite{Ell03,Hag02} and dark radiation\cite{Tan03,Nev03,Gum03,Yos04,Min04}.


 \section{Axial Vector Torsion and Gravity}

Cartan's structural equations for a Riemann-Cartan space-time $U_{4}$ are given by
\cite{Car22,Car24}
 \begin{eqnarray}T^{a}&=& de^{a}+\omega^{a}{}_{b}\wedge e^{b}\label{eqn:ab}\\
 R^{a}{}_{b}&=&d\omega^{a}{}_{b}+\omega^{a}{}_{c}\wedge \omega^{c}{}_{b},\label{eqn:ac}
\end{eqnarray}
here $\omega^{a}{}_{b}$ and e$^{a}$ represent the spin connection
and the local frame respectively.

In $U_{4}$ there exists  two invariant closed four forms. One is the well
known Pontryagin\cite{Che74,Che71} density \textit{P} and the
other is the less known Nieh-Yan\cite{Nie82} density \textit{N}
given by
\begin{eqnarray} \textit{P}&=& R^{ab}\wedge R_{ab}\label{eqn:ad}\\  \mbox{and} \hspace{2 mm}
 \textit{N}&=& d(e_{a}\wedge T^{a})\nonumber\\
&=&T^{a}\wedge T_{a}- R_{ab}\wedge e^{a}\wedge
e^{b}.\label{eqn:af}\end{eqnarray}

Here we consider a particular class of the Riemann-Cartan geometry where only the axial vector part of the torsion is nontrivial.
Then, from  (\ref{eqn:af}),  one naturally gets the Nieh-Yan
density 
\begin{eqnarray} N&=&-R_{ab}\wedge e^{a}\wedge e^{b}=-{}^* N\eta\hspace{2 mm},
\label{eqn:xaa}\\
 \mbox{where} \hspace{2 mm}\eta&:=&\frac{1}{4!}\epsilon_{abcd}e^{a}\wedge e^b\wedge
e^c\wedge e^d\end{eqnarray}is the invariant volume element.  It follows that    ${}^*N$,
the Hodge dual of $N$, is a scalar density of dimension $(length)^{-2}$.

We can combine the spin connection and the vierbeins multiplied by a scalar field together in a connection for $SO(4,1)$, in the tangent space, in the
form
\begin{eqnarray}W^{AB}&=&\left
[\begin{array}{cc}\omega^{ab}&\phi e^{a}\\- \phi e^{b}&0\end{array}\right],\label{eqn:aab}
\end{eqnarray}
where $a,b = 1,2,..4$; $A,B = 1,2,..5$ and $\phi$ is a variable
parameter of dimension $(length)^{-1}$ and Weyl weight $(-1)$, such
that, $\phi e^a$ has the correct  dimension and conformal weight
of the de Sitter boost part of the $SO(4,1)$ gauge connection. In
some earlier works\cite{Cha97,Mah02,Mah04}  $\phi$ has been
treated as an inverse length constant. In a recent
paper\cite{Mah05} $\phi$ has been associated, either in laboratory
scale or in cosmic sale, with a local energy scale. In laboratory
scale its coupling with torsion gives the mass term of the scalar
field and in cosmic scale it exactly produces the phenomenological
energy densities of the FRW universe.  The gravitational Lagrangian, in this approach, has been defined to be
\begin{eqnarray}\mbox{$\mathcal{L}$}_{G}&=& -\frac{1}{6}({}^*N \mbox{$\mathcal{R}$}\eta  +\beta \phi^2 N)+{}^*(b_a\wedge
\bar{\nabla} e^{a})(b_a\wedge
\bar{\nabla} e^{a}) \nonumber\\&{}&- f(\phi)d\phi\wedge{}^*d\phi -h(\phi)
\eta,\label{eqn:abcd1}\end{eqnarray} where
  *   is Hodge duality operator,    $\mathcal {R}$$\eta=\frac{1}{2}\bar{R}^{ab}\wedge\eta_{ab}$, $\bar{R}^b{}_a=d\bar{\omega}^b{}_a+\bar{\omega}^b{}_c\wedge \bar{\omega}^c{}_a$, $\bar{\omega}^{a}{}_{b}=\omega^{a}{}_{b}-T^{a}{}_{b}$, $ T^a=e^{a\mu}T_{\mu\nu\alpha}dx^\nu\wedge dx^\alpha$, $T^{ab}=e^{a\mu}e^{b\nu}T_{\mu\nu\alpha}  dx^\alpha$, $T=\frac{1}{3!}T_{\mu\nu\alpha}dx^\mu\wedge dx^\nu\wedge dx^\alpha$, $N=6dT$, $\eta_a=\frac{1}{3!}\epsilon_{abcd}e^b\wedge e^c\wedge e^d$ and $\eta_{ab}={}^*(e_a\wedge e_b)$. Here $\beta$ is a dimensionless coupling constant, $\bar{\nabla}$
represents covariant differentiation with respect to the connection one form $\bar{\omega}^{ab}$, $b_{  a}$ is a two form with
one internal index and of dimension $(length)^{-1}$ and $f(\phi)$,  $h(\phi)$ are unknown functions of $\phi$ whose
forms are to be determined subject to the geometric structure of the manifold.
  The geometrical implication of the first term, i.e. the $torsion \otimes curvature$\footnote{An important advantage of this part of the Lagrangian is that - it is a
   quadratic one with respect to the field derivatives and this
   could be valuable in relation to the quantization program of gravity like other gauge theories of
   QFT.} term, in the Lagrangian $\mathcal{L}\mbox{$_{G}$}$   has already been  discussed in the beginning.

  The Lagrangian $\mathcal{L}\mbox{$_{G}$}$
     is  only Lorentz invariant  under rotation in the tangent space where  de Sitter
     boosts are not permitted. As a consequence $T$ can be treated independently of $e^a$
     and $\bar{\omega}^{ab}$.
Here we note that, though torsion one form $T^{ab}=\omega^{ab}-\bar{\omega}^{ab}$ is a part of
the $SO(3,1)$ connection, it
 does not transform like a connection  form under $SO(3,1)$   rotation in
 the tangent space  and thus it imparts no constraint on the gauge degree of freedom of the
   Lagrangian.
 \section{Scalar Field and Spinorial Matter}

 Now we are in a position to write the total gravity Lagrangian in the presence of a spinorial matter field, given
by\begin{eqnarray}
\mbox{$\mathcal{L}$}_{tot.}&=&\mbox{$\mathcal{L}$}_{G}+\mbox{$\mathcal{L}$}_{D}, \label{eqn:apr}
\end{eqnarray} where
\begin{eqnarray}
\mbox{$\mathcal{L}$}_{D}&=&\phi^2[\frac{i}{2}\{\overline{\psi}{}^*\gamma\wedge D\psi+\overline{D\psi}\wedge{}^*\gamma\psi\}-\frac{g}{4}\overline{\psi}\gamma_5\gamma\psi\wedge T\nonumber\\&{}&+c_\psi\sqrt{{}^*dT} \overline{\psi}\psi\eta]\label{eqn:apr1}    \\
    \gamma_\mu &:=&\gamma_a e^a{}_\mu,\hspace{2mm}
    {}^*\gamma:=\gamma^a\eta_a,\hspace{2mm}
    D:=d+\Gamma\\
    \Gamma &:=&\frac{1}{4}\gamma^\mu D^{\{\}}\gamma_{\mu}=\frac{1}{4}\gamma^\mu \gamma_{\mu:\nu}dx^\nu\nonumber\\&=&-\frac{i}{4}\sigma_{ab}e^{a\mu}e^b{}_{\mu:\nu}dx^\nu
\end{eqnarray}
here $D^{\{\}}$, or $:$ in tensorial notation, is Riemannian torsion free covariant differentiation acting on external indices only; $\sigma^{ab}=\frac{i}{2}(\gamma^a\gamma^b-\gamma^b\gamma^a)$,  $\overline{\psi}=\psi^\dag\gamma^0$ and $g$, $c_\psi$ are both dimensionless coupling constants. Here $\psi$ and $\overline{\psi}$ have dimension $(length)^{-\frac{1}{2}}$ and conformal weight $-\frac{1}{2}$. It can be verified that under $SL(2,C)$ transformation on the spinor field and gamma matrices, given by,
\begin{eqnarray}    \psi\rightarrow\psi^\prime&=&S\psi,\hspace{2mm}\overline{\psi}\rightarrow\overline{\psi^\prime}=\overline{\psi}S^{-1}\nonumber\\\mbox{and}\hspace{2mm}\gamma\rightarrow\gamma^\prime&=&S\gamma S^{-1},
\end{eqnarray}where \mbox{$S=\exp(\frac{i}{4}\theta_{ab}\sigma^{ab})$,}
$\Gamma$ obeys the transformation property of a $SL(2,C)$ gauge connection, i.e.
\begin{eqnarray}
    \Gamma\rightarrow\Gamma^\prime&=&S(d+\Gamma)S^{-1}\\
    \mbox{s. t.}\hspace{2mm}D\gamma&:=&d\gamma+[\Gamma,\gamma]=0.\label{eqn:a47}
\end{eqnarray}Hence   $\gamma$ is a covariantly constant matrix valued one form w. r. t. the \mbox{$SL(2,C)$} covariant derivative $D$. By Geroch's theorem\cite{Ger68} we know that - the existence of  the spinor structure is equivalent to the existence of a global field of orthonormal tetrads on the space and time orientable manifold. Hence use of $\Gamma$ in the $SL(2,C)$ gauge covarint derivative is enough in a Lorentz invarint theory where de Sitter symmetry is broken.

In  appendices A and  B,  by varying  the independent fields in the Lagrangian $\mbox{$\mathcal{L}$}_{tot.}$, we obtain the Euler-Lagrange equations and then after some simplification we get the following results
\begin{flushright}$\begin{array}{lcr}
&\bar{\nabla} e_{a}=0,\hspace{38mm}&(\ref{eqn:abc9}^\prime)\nonumber\\&
{}^*N=\frac{6}{\kappa},\hspace{38mm}&(\ref{eqn:abc17}^\prime)\nonumber
\end{array}$\end{flushright}
 i.e. $\bar{\nabla}$ is torsion free and $\kappa$ is an integration constant having  dimension of $(length)^{2}$.\footnote{In (\ref{eqn:abcd1}), $\bar{\nabla}$ represents a $SO(3,1)$ covariant derivative, it is only on-shell torsion-free through the field equation (\ref{eqn:abc9}$^\prime$). The $SL(2,C)$ covariant derivative represented by the operator $D$ is torsion-free by definition, i.e. it is torsion-free both on on-shell and off-shell. Simultaneous and independent use of both $\bar{\nabla}$ and $D$ in the Lagrangian density (\ref{eqn:apr}) has been found to be advantageous in the approach of this article. This amounts to the emergence of the gravitational constant   $\kappa$ to be  only an on-shell  constant and this justifies the need for the introduction of the Lagrangian multiplier $b_a$ which appears twice in the Lagrangian density (\ref{eqn:abcd1}) such that $\bar{\omega}^a{}_b$ and $e^a$  become independent fields.}
\begin{flushright}$\begin{array}{lcr}   & m_\psi=c_\psi\sqrt{{}^*dT}=\frac{c_\psi}{\sqrt{\kappa}},&\hspace{20mm}(\ref{eqn:a59}^\prime)\nonumber
\\&i{}^*\gamma\wedge D\Psi-\frac{g}{4}\gamma_5\gamma\wedge T\Psi+m_\Psi\Psi\eta=0,&\nonumber\\&
i\overline{D\Psi}\wedge{}^*\gamma-\frac{g}{4}\overline{\Psi}\gamma_5\gamma\wedge T+m_\Psi\overline{\Psi}\eta=0,&\hspace{20mm} \nonumber  (\ref{eqn:a580}^\prime)
\end{array}$ \end{flushright} where $\Psi=\phi\psi$ and $ m_\Psi= m_\psi$.

\begin{flushright}$\begin{array}{rclr}
G^b{}_a\eta&=&-\kappa[ \frac{i}{8}\{\overline{\Psi}(\gamma^b D_a+\gamma_a D^b)\Psi-(\overline{D_a\Psi}\gamma^b+&\\&{}&\overline{D^b\Psi}\gamma_a)\Psi\}\eta-\frac{g}{16}\overline{\Psi}\gamma_5(\gamma_a {}^*T^b+\gamma^b {}^*T_a)\Psi\eta&\\&{}&+f\partial_a\phi\partial^b\phi\eta+\frac{1}{2}(h)\eta\delta^b{}_a],&(\ref{eqn:e1}^\prime)\\0&=&[\frac{1}{2}  \overline{\nabla}_\nu\overline{\Psi}\{\frac{\sigma^b{}_{a}}{2}, \gamma^\nu\} \Psi+\frac{i}{2}\{\overline{\Psi}(\gamma^b D_a-\gamma_a D^b)\Psi&\nonumber\\&{}&-(\overline{D_a\Psi}\gamma^b-\overline{D^b\Psi}\gamma_a)\Psi\} &\\&{}&-\frac{g}{4}\overline{\Psi}\gamma_5(\gamma_a {}^*T^b-\gamma^b {}^*T_a)\Psi]\eta,&(\ref{eqn:e2}^\prime)\nonumber
\end{array}$

$\begin{array}{lr} \kappa d[\frac{g}{4}{}^*(\overline{\Psi}\gamma_5\gamma\Psi\wedge T)-f{}^*(d\phi\wedge{}^*d\phi)+2h-\frac{\beta}{\kappa}\phi^2] &\\{}=-\frac{g}{4}\overline{\Psi}\gamma_5\gamma\Psi,&(\ref{eqn:abc141}^\prime)
    \\ \frac{2}{\kappa}\beta \phi + f^\prime(\phi)d\phi\wedge{}^*d\phi -h^\prime(\phi)
\eta+2fd{}^*d\phi&\\=-2\phi[\frac{i}{2}\{\overline{\psi}{}^*\gamma\wedge D\psi+\overline{D\psi}\wedge{}^*\gamma\psi\}&\\\hspace{4mm}-\frac{g}{4}\overline{\psi}\gamma_5\gamma\psi\wedge T+m_\psi \overline{\psi}\psi\eta]=0.&(\ref{eqn:abc142}^\prime)
\end{array}$ \end{flushright}

 Now let us make few comments about these results,
\begin{itemize}
\item Right hand side of equation (\ref{eqn:e1}$^\prime$) may be interpreted\cite{Bor02} as  ($-\kappa$) times the energy-momentum stress tensor of the Dirac field $\Psi(\overline{\Psi})$ together with the scalar field $\phi$. Where by equation
(\ref{eqn:abc17}$^\prime$) the gravitational constant $\kappa$ is
$\frac{6}{{}^*N}$ and then by equation (\ref{eqn:a59}$^\prime$) mass of the
spinor field is proportional to $\sqrt{{}^*N}$.
\item Equation (\ref{eqn:e2}$^\prime$)  represents covariant conservation of angular momentum of the Dirac field in the Einstein-Cartan space $U_4$ as a generalization of the same in the Minkwoski space $M_4$\cite{Itz85}.
\item Equation (\ref{eqn:abc142}$^\prime$) is the field equation of the scalar field $\phi$. Here it appears that, in the on-shell, other than gravity, it has no source. Whereas in equation (\ref{eqn:abc141}$^\prime$), there is a non trivial appearance of the torsion, the axial-vector matter-current and the scalar field $\phi$; provided the coupling constant $g$ is not negligible in a certain energy scale.
\end{itemize}
Since there is no direct experimental evidence of any torsion-matter interaction\cite{Sha01}, we may take,  $g$ to be  negligible at present time although there is a possibility that $g$ played a dominant role in the early universe. In other words, we may say that the scalar field, which appears to be connected with the spinor field only in equation (\ref{eqn:abc141}$^\prime$), is at present playing the role of the dark matter and/or dark radiation.   The consequence of this  spin-torsion interaction term, in the very early universe,   may be linked to the cosmological inflation without false vacuum\cite{Gas86}, primordial density fluctuation \cite{And99a,Pal99} and/or to the repulsive gravity\cite{Gas98}.
 \section{Scalar Field, Dark Matter and Da- rk Radiation}
Now let us analyse our results in the background of a FRW-cosmology
where the metric tensor is given by
\begin{eqnarray}
 g_{00}=-1 ,\hspace{2mm}g_{ij}= \delta_{ij}a^2(t)  \hspace{2mm}\mbox{where}\hspace{2mm}
i,j=1,2,3;\label{eqn:aay}
\end{eqnarray}such that
\begin{eqnarray}
    &{}&e=\sqrt{-\det(g_{\mu\nu})}= a^3\label{eqn:zaz}
\end{eqnarray}Taking $g=0$, equation (\ref{eqn:abc141}$^\prime$) gives us  \begin{eqnarray}
     f \dot{\phi}^2=-\frac{1}{\kappa}(\beta\phi^2 +\lambda)+2h .\label{eqn:aaj}
\end{eqnarray}where $\lambda$ is a constant of integration of dimension $(length)^{-2}$.
   Now, with the cosmological restriction on the metric  as stated in  (\ref{eqn:aay})
 and the $\phi$-field is a function of time only, the equation (\ref{eqn:abc142}$^\prime$) reduces to
\begin{eqnarray}
     2f \ddot{\phi}+2f\frac{e^\prime}{e}\dot{\phi}^2+ f^\prime\dot{\phi}^2-\frac{2\beta}
{\kappa}\phi + h^\prime=0
\end{eqnarray}where ${}^\prime$ represents differentiation w. r. t. $\phi$. If we eliminate $\ddot{\phi}$ from this equation with the help of the time derivative of equation
(\ref{eqn:aaj}),  we get
\begin{eqnarray}
 2f\frac{e^\prime}{e}\dot{\phi}^2&=&\frac{4\beta}{\kappa}\phi-3h^\prime \nonumber\\
 \mbox{or,}\hspace{5mm} 2\frac{e^\prime}{e}&=&-\frac{\frac{4\beta}{\kappa}\phi-3h^\prime}{\frac{1}{\kappa}(\beta\phi^2 +\lambda)-2h}\label{eqn:aal}
\end{eqnarray}

Now, for the FRW metric, the non-vanishing components of Einstein's tensor (\ref{eqn:e1}$^\prime$), w. r. t. external indices,
are given by
\begin{eqnarray}
    G^{0}{}_{0}&=&-3(\frac{\dot{a}}{a})^2=-\kappa(\rho_{BM}+\frac{\beta}{\kappa}\phi^{2}+\frac{\lambda}
{\kappa}-\frac{3h}{2}) \nonumber\\G^{j}{}_{i}&=&-(\frac{2\ddot{a}}{a}+\frac{\dot{a}^2}{a^2})
\delta^{j}{}_{i}=-\kappa\frac{1}{2}(h)\delta^{j}{}_{i}\label{eqn:11a}
\end{eqnarray}where we have assumed that, in the cosmic scale, the observed (luminous) mass distribution is baryonic and co moving, s. t.
\begin{eqnarray}
    &{}&\mathop\sum\limits_{\Psi}^{}\frac{i}{8}\{\overline{\Psi}(\gamma^b D_a+\gamma_a D^b)\Psi-(\overline{D_a\Psi}\gamma^b+\overline{D^b\Psi}\gamma_a)\Psi\}\nonumber\\&{}&\hspace{6mm}=\rho_{BM}=\frac{M_{BM}}{V},\hspace{2mm}\mbox{for $a=b=0$},\nonumber\\&{}&\hspace{6mm}=0,\hspace{2mm}\mbox{otherwise}.
\end{eqnarray}Here $M_{BM}$ and $V$ are the total baryonic mass and volume of the universe respectively.

From the forms of $G^{0}{}_{0}$ and $G^{j}{}_{i}$ it appears that the term $\frac{\beta}
{\kappa}\phi^{2}$ represents pressure-less energy density i.e. $\phi^{2}\propto a^{-3}\propto
 \frac{1}{e}$. Putting this in (\ref{eqn:aal}) we get after integration
\begin{eqnarray}
    h=-\gamma \phi^{\frac{8}{3}}+\frac{\lambda}{2\kappa}
\end{eqnarray} where $\gamma$ is a constant of dimension $(length)^{-\frac{4}{3}}$ .
Then from  (\ref{eqn:11a}), we get
\begin{eqnarray}
G^{0}{}_{0}&=&-3(\frac{\dot{a}}{a})^2=-\kappa(\rho_{BM}+\rho_{DM}+\rho_{DR}+\rho_{VAC.}) \nonumber\\G^{j}{}_{i}&=&-(\frac{2\ddot{a}}{a}+\frac{\dot{a}^2}{a^2})
\delta^{j}{}_{i}\nonumber\\&=&\kappa(p_{BM}+p_{DM}+p_{DR}+p_{VAC.}) \label{eqn:11b}\delta^{j}{}_{i},
\end{eqnarray}  where
\begin{eqnarray}
    p_{BM}=p_{DM}=0\\
    \rho_{DM}=\frac{\beta}{\kappa}\phi^2\\
    \rho_{DR}=\frac{3\gamma}{2}\phi^{\frac{8}{3}},\hspace{2mm}p_{DR}=\frac{1}{3}\rho_R\\
\rho_{VAC.}=-p_{VAC.}=\frac{\lambda}{4\kappa}=\Lambda \mbox{ (say)}.
\end{eqnarray}As the scalar field $\phi$, at present scale, appears to be non-inter- acting with the spinor field $\Psi$, \textit{vide equations} (\ref{eqn:a580}$^\prime$), (\ref{eqn:abc141}$^\prime$) \& (\ref{eqn:abc142}$^\prime$), the quantities having subscripts ${}_{BM}$, ${}_{DM}$,   ${}_{DR}$ and ${}_{VAC.}$ may be assigned to the baryonic matter, the dark matter, the dark radiation and the vacuum energy respectively. If we add another Lagrangian density to (\ref{eqn:apr1}) corresponding to the Electro-Magnetic field and modify $D$ by $D+A$, where $A$ is the $U(1)$   connection one form, and also consider massless spinors having $c_\psi=0$, then on the r. h. s. of equations in (\ref{eqn:11b}),  $\rho_{DR}$ and $p_{DR}$ would be replaced by $\rho_R$ and $p_{R}$  containing  various  radiation components, given by
\begin{eqnarray}
    \rho_R&=&\rho_{DR}+\rho_\gamma+\rho_\nu ,\nonumber\\
    p_R&=&p_{DR}+p_\gamma+p_\nu\\
    \mbox{s. t.}\hspace{2mm}p_R&=&\frac{1}{3}\rho_R ,
\end{eqnarray}where the subscripts have their usual meanings.
Then from (\ref{eqn:aaj}) and (\ref{eqn:11b}), we get
\begin{eqnarray}
    f=-\frac{1}{\kappa\phi^2\rho}(\frac{8}{3}\rho_{DR}+\frac{4}{3}\rho_{DM}),\label{eqn:111a}
\end{eqnarray}where $\rho$ may  be written as
\begin{eqnarray}
    \rho=(1+\frac{\rho_{BM}}{\rho_{DM}})\rho_{DM}+(1+\frac{\rho_\gamma}{\rho_{DR}}+\frac{\rho_\nu}{\rho_{DR}})\rho_{DR}+\rho_{VAC.} .\label{eqn:11c}
\end{eqnarray}Here the dimensionless parameter $\frac{\rho_{BM}}{\rho_{DM}}$ is the baryonic matter-dark matter ratio of the universe and, as both $\rho_{BM}$ \& $\rho_{DM}$ have the same power-law of evolution at large cosmic scale, it  may be taken to be a   constant of time. Similarly  the parameters $\frac{\rho_\gamma}{\rho_{DR}}$ and $\frac{\rho_\nu}{\rho_{DR}}$ may also be taken to be constants of time. And then from (\ref{eqn:111a}), $f$ may be expressed in the following form,
\begin{eqnarray}
    f=-\frac{A+B\phi^{\frac{2}{3}}}{C\phi^2+D\phi^{\frac{8}{3}}+E}\label{eqn:111c},
\end{eqnarray}where $A$, $B$, $C$, $D$ and $E$ are   constants having proper dimensions. It may be checked that $f\propto\phi^{-2}$(approx.) in both matter and radiation dominated era of the universe but $f$ is nearly a  constant at a very late time when the energy density is dominated by the cosmological constant.
\section{Discussion} In this article, we have seen that if we introduce a scalar field $\phi$ to cause  the de Sitter connection to have the proper dimension of a gauge field and also link this scalar field with the dimension of a Dirac field then we find that the Euler-Lagrange equations of both the fields to be mutually non-interacting. But they are indirectly connected to each other when we consider Euler-Lagrange equations of other geometric fields such as torsion and tetrad.  Variation of the $SO(3,1)$ spin connection as an entity independent of the tetrads we get the Newton's constant as inversely proportional to the topological Nieh-Yan density  and then the mass of the spinor field has been shown to be linked to the Newton's constant. Then using symmetries of the Einstein's tensor we get covariant conservation of angular momentum of the Dirac field in the particular class of geometry in  $U_4$ as a generalization of the same in the Minkwoski space $M_4$.

In the present scale of the universe we may neglect the  spin-torsion interaction term and then considering FRW cosmology we are able to derive standard cosmology with standard energy density together with dark matter, dark radiation and cosmological constant. From brane cosmology we know that dark radiation should strongly affect both the big-bang nucleosynthesis(BBN) and the cosmic microwave background(CMB) and, in particular, constraints on BBN alone allow dark radiation to be significant\cite{Ich02}. Therefore appearance of dark radiation from a different point of view, as pursued in this article, is very significant. In this present analysis it is significant that if we consider our universe to have the isotropy and the homogeneity of a FRW universe then only three kinds of energy densities are possible. The matter energy density $\propto a^{-3}$, the radiation energy density $\propto a^{-4}$ and the vacuum energy density $\propto a^0$ are the only three kinds of such energy densities where $a$ is the cosmic scale factor. It is surprising that these are the only three kinds of phenomenological cosmic energy density that we observe and consider to be interested in. But theoretically, in standard FRW cosmology, other forms of energy density are not ruled out\cite{Mah05}. And therefore to consider other forms of cosmic energy density, may be in the early universe, we have to adopt a non-FRW geometry where we may have to forgo the isotropy and the homogeneity of the universe. There we may  have to invite the spin-torsion interaction term to play a dominant role. The consequence of the  spin-torsion interaction term, in the very early universe,   may be linked to the cosmological inflation without false vacuum\cite{Gas86}, primordial density fluctuation \cite{And99a,Pal99} and/or to the repulsive gravity\cite{Gas98}. It is important to note that, in a model,   Capozziello et al\cite{Cap01} have considered    a totally antisymmetric torsion field to
discuss  the conditions for quintessence and have obtained exact
solutions   where dust dominated Friedmann
behavior have been  recovered as soon as torsion effects become not relevant. In this context we may mention a recent work\cite{Mah02} where, in the gravity without metric formalism of gravity, a particular canonical transformation of the field variables causes the simultaneous appearance of the CP-violating $\theta$-term and the cosmological term. Here we also like to consider  the finding of
some other works, where the gauge  group is   $SL(2,C)$  , where torsion has been shown to be
linked with CP-violation\cite{Ban95} and  fermion mass\cite{Ban00}.
In a recent work\cite{Kir05,Kir06}, using a nonlinear realization technique of the algebra of diffeomorphism, a Higgs mechanism of gravity has been constructed in which an affine space time evolves into a Riemannian one by the condensation of a metric where the symmetry breaking potential is identical to that of hybrid inflation. In this approach torsion can be recovered if the space time coordinates are allowed to be non-commutative.  In another work\cite{Lan01}, it has been shown that torsion is a natural consequence in non-commutative $U(1)$ Yang-Mills theory where gauge symmetries give a very natural and explicit realizations of the mixing of space-time and internal symmetries. Here torsion measures the non-commutativity of displacement of points in the flat space-time in the teleparallel theory and the non-commutativity scale is given by the Planck length.
 As a final comment we may say that - the mutual role of gravity and matter
 becomes transparent only when we consider that the curvature and the torsion play complementary roles in the geometry.
 \section*{Acknowledgment}
           I wish to thank Prof. Pratul Bandyopadhyay, Indian Statistical Institute,  Kolkata,  for his valuable remarks and fruitful suggestions on this
           problem.

\renewcommand{\theequation}{\mbox{A}\arabic{equation}}
\setcounter{equation}{0}

\begin{appendix}
\section*{Appendix A}
 Following reference \cite{Heh95}, we independently vary
     $e^{a}$, $\bar{\nabla} e^{a}$, $dT$,  $\bar{R}^{ab}$, $\phi$, $d\phi$ and
    $b^{a}$  and find

\begin{eqnarray}
    \delta  \mathcal{L}\mbox{$_{G}$}&=&\delta e^a\wedge \frac{\partial
    \mathcal{L}\mbox{$_{G}$}}{\partial e^a}+\delta \bar{\nabla} e^a\wedge \frac{\partial
    \mathcal{L}\mbox{$_{G}$}}{\partial \bar{\nabla} e^a}+\delta dT \frac{\partial  \mathcal{L}\mbox{$_{G}$}}{\partial
    dT}\nonumber\\&{}&+\delta \bar{R}^{ab}\wedge \frac{\partial  \mathcal{L}
\mbox{$_{G}$}}{\partial
    \bar{R}^{ab}}+\delta\phi\frac{\partial  \mathcal{L}\mbox{$_{G}$}}{\partial \phi}+\delta
    d\phi\wedge\frac{\partial  \mathcal{L}\mbox{$_{G}$}}{\partial d\phi}\nonumber\\&{}&+\delta b^a\wedge
    \frac{\partial  \mathcal{L}\mbox{$_{G}$}}{\partial b^a} \\&=&\delta e^a\wedge( \frac{\partial
    \mathcal{L}\mbox{$_{G}$}}{\partial e^a}+ \bar{\nabla}\frac{\partial  \mathcal{L}
    \mbox{$_{G}$}}{\partial \bar{\nabla} e^a})+\delta T\wedge d \frac{\partial  \mathcal{L}
    \mbox{$_{G}$}}{\partial dT}\nonumber\\&{}&+\delta \bar{\omega}^{ab}\wedge(\bar{\nabla} \frac{\partial
     \mathcal{L}\mbox{$_{G}$}}{\partial \bar{R}^{ab}}+ \frac{\partial
    \mathcal{L}\mbox{$_{G}$}}{\partial \bar{\nabla} e^a}\wedge e_b)\nonumber\\&{}&+\delta \phi(\frac{\partial
    \mathcal{L}\mbox{$_{G}$}}{\partial \phi}- d \frac{\partial  \mathcal{L}\mbox{$_{G}$}}
    {\partial d\phi})+\delta b^a\wedge \frac{\partial  \mathcal{L}\mbox{$_{G}$}}{\partial
    b^a}\nonumber \\&{}&
    +d(\delta e^a\wedge \frac{\partial  \mathcal{L}\mbox{$_{G}$}}{\partial
    \bar{\nabla} e^a}+\delta T \frac{\partial  \mathcal{L}\mbox{$_{G}$}}{\partial dT}\nonumber\\&{}&+\delta
    \bar{\omega}^{ab}\wedge \frac{\partial  \mathcal{L}\mbox{$_{G}$}}{\partial \bar{R}^{ab}}+
    \delta\phi\frac{\partial  \mathcal{L}\mbox{$_{G}$}}{\partial d\phi})\label{eqn:abc0}
\end{eqnarray}Using the form of the Lagrangian $\mathcal{L}\mbox{$_{G}$}$, given in (\ref{eqn:abcd1}),
we get
\begin{eqnarray}
    \frac{\partial  \mathcal{L}\mbox{$_{G}$}}{\partial e^a}&=&-\frac{1}{6}{}^*N(2\textbf{R}_a-\mbox{$\mathcal{R}$}\eta_a)
  -{}^*(b_b\wedge
\bar{\nabla} e^{b})^2\eta_a\nonumber\\&{}&-f(\phi)[-2\partial_a\phi\partial^b\phi\eta_b+\partial_b\phi\partial^b\phi\eta_a]\nonumber\\&{}&-h(\phi)\eta_a\label{eqn:abc1} \\
\frac{\partial  \mathcal{L} \mbox{$_{G}$}}{\partial (\bar{\nabla}
e^a)}&=&2{}^*(b_a\wedge \bar{\nabla} e^{a})b_a\label{eqn:abc2}
 \\\frac{\partial  \mathcal{L}\mbox{$_{G}$}}{\partial (dT)}&=&\mbox{$\mathcal{R}$}-\beta\phi^2\label{eqn:abc4}\\\frac{\partial  \mathcal{L}\mbox{$_{G}$}}{\partial \bar{R}^{ab}}&=&-\frac{1}{24}{}^*N\epsilon_{abcd}e^c\wedge e^d=-\frac{1}{12}{}^*N\eta_{ab}\label{eqn:abc5}\\\frac{\partial  \mathcal{L}\mbox{$_{G}$}}{\partial \phi}&=&-\frac{1}{3}\beta \phi N- f^\prime(\phi)d\phi\wedge{}^*d\phi -h^\prime(\phi)
\eta\label{eqn:abc5a}\\\frac{\partial  \mathcal{L}\mbox{$_{G}$}}{\partial d\phi}&=&-2f{}^*d\phi\label{eqn:abc5b}\\\frac{\partial  \mathcal{L}\mbox{$_{G}$}}{\partial b^a}&=&2{}^*(b_b\wedge
\bar{\nabla} e^{b})
\bar{\nabla} e_{a}\label{eqn:abc6}
\eea
Where
\bea
     \textbf{R}_a:=\frac{1}{2}\frac{\partial (\mbox{$\mathcal{R}$}\eta)}{\partial e^a}&=&\frac{1}{4}\epsilon_{abcd}\bar{R}^{bc}\wedge  e^d\hspace{2mm}\nonumber\\\mbox{and}\hspace{2mm}\partial_a&=&e_a{}^\mu\frac{\partial}{\partial x^\mu} .
\end{eqnarray}

From above, Euler-Lagrange equations for  $b_a$ gives us
\begin{eqnarray}
\bar{\nabla} e_{a}&=&0\label{eqn:abc9}
\end{eqnarray}i.e.   $\bar{\nabla}$ is torsion free. Using this result in (\ref{eqn:abc1}) and  (\ref{eqn:abc2})  we get
\begin{eqnarray}
    \frac{\partial  \mathcal{L}\mbox{$_{G}$}}{\partial e^a}&=&-\frac{1}{6}{}^*N(2\textbf{R}_a-\mbox{$\mathcal{R}$}\eta_a)-f(\phi)[-2\partial_a\phi\partial^b\phi\eta_b\nonumber\\&{}&+\partial_b\phi\partial^b\phi\eta_a]-h(\phi)\eta_a\label{eqn:abc10}\\    \frac{\partial  \mathcal{L}\mbox{$_{G}$}}{\partial (\bar{\nabla} e^a)}&=&0\label{eqn:abc11}
\end{eqnarray}
\renewcommand{\theequation}{\mbox{B}\arabic{equation}}
\setcounter{equation}{0}
\section*{Appendix B}Independent variation of $\psi$, $\overline{\psi}$, $D\psi$, $\overline{D\psi}$, $T$, $dT$, $\phi$ and $e^a$ in $\mathcal{L}$$_{D}$ gives us
\begin{eqnarray}    \delta\mbox{$\mathcal{L}$}_{D}&=&\frac{\partial\mbox{$\mathcal{L}$}_{D}}{\partial\psi}\delta \psi+\frac{\partial\mbox{$\mathcal{L}$}_{D}}{\partial (D\psi)}\wedge\delta (D\psi)+\delta \overline{\psi}\frac{\partial\mbox{$\mathcal{L}$}_{D}}{\partial\overline{\psi}}\nonumber\\&{}&+\delta \overline{D\psi}\wedge\frac{\partial\mbox{$\mathcal{L}$}_{D}}{\partial \overline{D\psi}}+\delta T\wedge\frac{\partial\mbox{$\mathcal{L}$}_{D}}{\partial T}+\delta (dT)\frac{\partial\mbox{$\mathcal{L}$}_{D}}{\partial (dT)}\nonumber\\&{}&+\delta \phi \frac{\partial\mbox{$\mathcal{L}$}_{D}}{\partial \phi}+\delta e^a\wedge\frac{\partial\mbox{$\mathcal{L}$}_{D}}{\partial e^a}\label{eqn:a53}
\end{eqnarray}
Now $\delta(D\psi)=D\delta\psi+\delta\Gamma\psi$, since $d$ and $\delta$ commute, reduces (\ref{eqn:a53}) to
\begin{eqnarray}
    \delta\mbox{$\mathcal{L}$}_{D}&=& \phi^2[\{i\overline{D\psi}\wedge{}^*\gamma+id\ln\phi\wedge\overline{\psi}{}^*\gamma-\frac{g}{4}\overline{\psi}\gamma_5\gamma\wedge T\nonumber\\&{}&+c_\psi\sqrt{{}^*dT}\overline{\psi}\eta\}\delta\psi+\delta\overline{\psi}\{i{}^*\gamma\wedge D\psi\nonumber\\&{}&+i{}^*\gamma\psi\wedge d\ln\phi-\frac{g}{4}\gamma_5\gamma\wedge T\psi+c_\psi\sqrt{{}^*dT}\psi\eta\}\nonumber\\&{}&+\frac{i}{2}\{\overline{\delta\Gamma\psi}\wedge{}^*\gamma\psi+\overline{\psi}{}^*\gamma\wedge\delta\Gamma\psi\}\nonumber\\&{}&+\frac{i}{2}\{\overline{\psi}\delta({}^*\gamma)\wedge D\psi+\overline{D\psi}\wedge\delta({}^*\gamma)\psi\}\nonumber\\&{}&+\delta T\wedge\{\frac{g}{4}\overline{\psi}\gamma_5\gamma\psi-\frac{1}{2\phi^2}c_\psi d(\frac{\phi^2}{\sqrt{{}^*dT}}\overline{\psi} \psi)\}\nonumber\\&{}&+\delta e^a\wedge\{-\frac{g}{4}\overline{\psi}\gamma_5\gamma_a\psi T+\frac{1}{2}c_\psi\sqrt{{}^*dT} \overline{\psi}\psi\eta_a\}]   \nonumber\\
    &{}&+2\phi[\frac{i}{2}\{\overline{\psi}{}^*\gamma\wedge D\psi+\overline{D\psi}\wedge{}^*\gamma\psi\}\nonumber\\&{}&-\frac{g}{4}\overline{\psi}\gamma_5\gamma\psi\wedge T+c_\psi\sqrt{{}^*dT} \overline{\psi}\psi\eta]\delta\phi\nonumber\\&{}&+\hspace{2mm}\mbox{surface terms(S. T.)}.\label{eqn:a54}
\end{eqnarray} Third term of this equation gives
\begin{eqnarray}    &{}&\frac{i}{2}[\overline{\delta\Gamma\psi}\wedge{}^*\gamma\psi+\overline{\psi}{}^*\gamma\wedge\delta\Gamma\psi]\phi^2\nonumber\\&{}&=\frac{i}{2}\overline{\psi}[-\delta\Gamma\wedge{}^*\gamma+{}^*\gamma\wedge\delta\Gamma]\psi\phi^2\nonumber\\&{}&=-\frac{1}{8}\overline{\psi}[\sigma_{cb}\delta(e^{c\mu}e^b{}_{\mu:\nu})\gamma^\nu+\gamma^\nu\sigma_{cb}\delta(e^{c\mu}e^b{}_{\mu:\nu})]\psi\phi^2\eta\nonumber\\&{}&=\frac{1}{8}\delta e^a{}_\alpha[\overline{\psi}(\sigma_{cb}e^{c\alpha}e_a{}^\mu e^b{}_{\mu:\nu}\gamma^\nu+\gamma^\nu\sigma_{cb}e^{c\alpha}e_a{}^\mu e^b{}_{\mu:\nu})\psi\phi^2\eta\nonumber\\&{}&\hspace{4mm}+D^{\{\}}_\nu\{\overline{\psi}(\sigma_{ca}e^{c\alpha} \gamma^\nu+\gamma^\nu\sigma_{ca}e^{c\alpha} )\psi\phi^2\eta\}]+\hspace{2mm}\mbox{(S. T.)}\nonumber\\&{}&=\frac{1}{8}\delta e^a\wedge[\overline{\psi}(\sigma_{cb} e_a{}^\mu e^b{}_{\mu:\nu}\gamma^\nu+\gamma^\nu\sigma_{cb} e_a{}^\mu e^b{}_{\mu:\nu})\psi\phi^2\eta^c\nonumber\\&{}&\hspace{4mm}+D^{\{\}}_\nu\{\overline{\psi}(\sigma_{ca} \gamma^\nu+\gamma^\nu\sigma_{ca} )\psi\phi^2\eta^c\}]+\hspace{2mm}\mbox{(S. T.)}\label{eqn:a55}
    \end{eqnarray}It is to be noted here that, using the properties of $\gamma$ matrices, the only surviving terms of
         $\frac{i}{2}[\overline{\Gamma\psi}\wedge{}^*\gamma\psi+\overline{\psi}{}^*\gamma\wedge\Gamma\psi]$ $=$ $-\frac{i}{2}\overline{\psi}(\gamma^\mu\gamma_{\mu:\nu}\gamma^\nu+\gamma^\nu\gamma^\mu\gamma_{\mu:\nu})\psi \eta$ are those for which $\gamma^\mu$, $\gamma^\nu$ and $\gamma_{\mu:\nu}$ are anti-symmetrized. This implies that, in the variational calculation, the Christoffel part of the Riemannian covariant derivative remains insensitive.

    Fourth term of (\ref{eqn:a54}) gives
\begin{eqnarray}&{}&
\frac{i}{2}\{\overline{\psi}\delta({}^*\gamma)\wedge D\psi+\overline{D\psi}\wedge\delta({}^*\gamma)\psi\}\phi^2 \nonumber\\&{}&=\frac{i}{2}\delta e^a\wedge \{\overline{\psi}\gamma^b\eta_{ba}\wedge D\psi-\overline{D\psi}\gamma^b\psi\wedge\eta_{ba}\}\phi^2\label{eqn:a56}
\end{eqnarray}

Hence Euler-Lagrange equations corresponding to the  extremum of $\mbox{$\mathcal{L}$}_{tot.}$ from the independent variations  of $e^a$, $T$, $\phi$ and $\bar{\omega}^{ab}$, using (\ref{eqn:abc0}),  (\ref{eqn:abc4})  and (\ref{eqn:abc5}), give us
\begin{eqnarray}&{}&\frac{1}{6}    {}^*N(2\textbf{R}_a-\mbox{$\mathcal{R}$}\eta_a)\nonumber\\&{}&+f(\phi)[-2\partial_a\phi\partial^b\phi\eta_b+\partial_b\phi\partial^b\phi\eta_a]+h(\phi)\eta_a\nonumber\\&{}&-\frac{1}{8}[\overline{\psi}(\sigma_{cb} e_a{}^\mu e^b{}_{\mu:\nu}\gamma^\nu+\gamma^\nu\sigma_{cb} e_a{}^\mu e^b{}_{\mu:\nu})\psi\phi^2\eta^c\nonumber\\&{}&+D^{\{\}}_\nu\{\overline{\psi}(\sigma_{ca} \gamma^\nu+\gamma^\nu\sigma_{ca} )\psi\phi^2\eta^c\}]\nonumber\\&{}&+[\frac{g}{4}\overline{\psi}\gamma_5\gamma_a\psi\wedge T-\frac{1}{2}c_\psi\sqrt{{}^*dT} \overline{\psi}\psi\eta_a]\phi^2\nonumber\\&{}&-\frac{i}{2} \{\overline{\psi}\gamma^b\eta_{ba}\wedge D\psi-\overline{D\psi}\gamma^b\psi\wedge\eta_{ba}\}\phi^2=0\label{eqn:abc13}
\end{eqnarray}
\begin{eqnarray}&{}&
d(\mbox{$\mathcal{R}$}-\beta\phi^2-c_\psi \frac{\phi^2}{2\sqrt{{}^*dT}}\overline{\psi} \psi)=-\frac{g}{4}\phi^2\overline{\psi}\gamma_5\gamma\psi\label{eqn:abc14}
\\&{}&-2\beta \phi N+ f^\prime(\phi)d\phi\wedge{}^*d\phi -h^\prime(\phi)
\eta+2fd{}^*d\phi\nonumber\\&{}&=-2\phi[\frac{i}{2}\{\overline{\psi}{}^*\gamma\wedge D\psi+\overline{D\psi}\wedge{}^*\gamma\psi\}\nonumber\\&{}&\hspace{4mm}-\frac{g}{4}\overline{\psi}\gamma_5\gamma\psi\wedge T+c_\psi\sqrt{{}^*dT} \overline{\psi}\psi\eta]\label{eqn:abc140}
\end{eqnarray}
\begin{eqnarray}
\bar{\nabla}({}^*N\eta_{ab})&=&0\label{eqn:abc15}
\end{eqnarray}
Using (\ref{eqn:abc9}) in (\ref{eqn:abc15}), we get
\begin{eqnarray}
    d{}^*N=0\label{eqn:abc16}
\end{eqnarray}
From  this equation  we can write
\begin{eqnarray}
{}^*N=\frac{6}{\kappa}\label{eqn:abc17}
\end{eqnarray}where $\kappa$ is an integration constant having $(length)^{2}$ dimension.
From (\ref{eqn:a54}) we can write the Euler-Lagrange equations for the fields $\psi$ and $\overline{\psi}$ as
\begin{eqnarray}
&{}&i{}^*\gamma\wedge D\psi+i{}^*\gamma\psi\wedge d\ln\phi\nonumber\\&{}&-\frac{g}{4}\gamma_5\gamma\wedge T\psi+m_\psi\psi\eta=0,\nonumber\\&{}&
i\overline{D\psi}\wedge{}^*\gamma+id\ln\phi\wedge\overline{\psi}{}^*\gamma\nonumber\\&{}&-\frac{g}{4}\overline{\psi}\gamma_5\gamma\wedge T+m_\psi\overline{\psi}\eta=0,  \label{eqn:a58}
\end{eqnarray}provided we define, using (\ref{eqn:abc17}), the mass of the field $\psi$ as
\begin{eqnarray}
    m_\psi=c_\psi\sqrt{{}^*dT}=\frac{c_\psi}{\sqrt{\kappa}}\label{eqn:a59}
\end{eqnarray}  If we define $\Psi=\phi\psi$ as the Dirac field having the proper dimension and conformal weight and $m_\Psi=m_\psi$, the equations in (\ref{eqn:a58}) reduce to their standard form in the particular class of geometry in $U_4$ space\cite{Mie01}
\bea
i{}^*\gamma\wedge D\Psi-\frac{g}{4}\gamma_5\gamma\wedge T\Psi+m_\Psi\Psi\eta&=&0,\nonumber\\
i\overline{D\Psi}\wedge{}^*\gamma-\frac{g}{4}\overline{\Psi}\gamma_5\gamma\wedge T+m_\Psi\overline{\Psi}\eta&=&0.   \label{eqn:a580}
\eea

Now using (\ref{eqn:abc17}), (\ref{eqn:a59}) and (\ref{eqn:a580}), the field equation (\ref{eqn:abc13}) reduces to
\begin{eqnarray}
&{}&\frac{1}{\kappa}    (2\textbf{R}_a-\mbox{$\mathcal{R}$}\eta_a)+f(\phi)[-2\partial_a\phi\partial^b\phi\eta_b+\partial_b\phi\partial^b\phi\eta_a]\nonumber\\&{}&+h(\phi)\eta_a-\frac{1}{8}\overline{\nabla}_\nu\{\overline{\Psi}(\sigma_{ca} \gamma^\nu+\gamma^\nu\sigma_{ca} )\Psi\eta^c\}\nonumber\\&{}&+[\frac{g}{4}\overline{\Psi}\gamma_5\gamma_a\Psi\wedge T-\frac{1}{2}m_\Psi \overline{\Psi}\Psi\eta_a]\nonumber\\&{}&-\frac{i}{2}    \{\overline{\Psi}\gamma^b\eta_{ba}\wedge D\Psi-\overline{D\Psi}\gamma^b\Psi\wedge\eta_{ba}\}=0\label{eqn:abc131}
\end{eqnarray}
where $\overline{\nabla}_\nu$ represents  torsion-free covariant differentiation w. r. t. both external and internal indices.
Now by exterior multiplication with $e^a$ from left, this equation yields
\begin{eqnarray}    &{}&\frac{2}{\kappa}\mbox{$\mathcal{R}$}\eta+\frac{3i}{2}\{\overline{\Psi} {}^*\gamma\wedge D\Psi+\overline{D\Psi}\wedge{}^*\gamma\Psi\}-\frac{g}{4}\overline{\Psi}\gamma_5\gamma\Psi\wedge T\nonumber\\&{}&+2m_\Psi  \overline{\Psi}\Psi\eta-2fd\phi\wedge{}^*d\phi-4h\eta=0\label{eqn:a62}
\end{eqnarray}and after using Dirac equations (\ref{eqn:a580}), we get
\begin{eqnarray}    \mbox{$\mathcal{R}$}\eta&=&\kappa[\frac{1}{2}m_\Psi\overline{\Psi}\Psi\eta-\frac{g}{4}\overline{\Psi}\gamma_5\gamma\Psi\wedge T\nonumber\\&{}&+fd\phi\wedge{}^*d\phi+2h\eta]\label{eqn:a63}
    \end{eqnarray}
Again, taking exterior multiplication of (\ref{eqn:abc131}) by $e^b$, we get
\begin{eqnarray}
&{}&\frac{1}{\kappa}[\mbox{$\mathcal{R}$}\delta^b{}_a+2G^b{}_a]\eta+\frac{i}{2} \{\overline{\Psi} {}^*\gamma\wedge D\Psi+\overline{D\Psi}\wedge{}^*\gamma\Psi\}\delta^b{}_a\nonumber\\&{}&+\frac{i}{2}\{\overline{\Psi}\gamma^bD_a\Psi-\overline{D_a\Psi}\gamma^b\Psi\}\eta-\frac{g}{4}\overline{\Psi}\gamma_5\gamma_a\Psi {}^*T^b\eta\nonumber\\&{}&+\frac{1}{2}m_\Psi\overline{\Psi}\Psi\delta^b{}_a\eta+\frac{1}{8}  \overline{\nabla}_\nu\{\overline{\Psi}(\sigma^b{}_{a} \gamma^\nu+\gamma^\nu\sigma^b{}_{a} )\Psi\}\eta\nonumber\\&{}&+2f\partial_a\phi\partial^b\phi\eta-fd\phi\wedge{}^*d\phi\delta^b{}_a-h\eta\delta^b{}_a=0  \label{eqn:a67}
\end{eqnarray}Using Dirac equations (\ref{eqn:a580}) and equation (\ref{eqn:a63}), this equation reduces to
\begin{eqnarray}
G^b{}_a\eta&=&-\kappa[ \frac{i}{4}\{\overline{\Psi}\gamma^b D_a\Psi-\overline{D_a\Psi}\gamma^b\Psi\}\eta-\frac{g}{8}\overline{\Psi}\gamma_5\gamma_a\Psi {}^*T^b\eta\nonumber\\&{}&+\frac{1}{16}  \overline{\nabla}_\nu\{\overline{\Psi}(\sigma^b{}_{a} \gamma^\nu+\gamma^\nu\sigma^b{}_{a} )\Psi\}\eta\nonumber\\&{}&+f\partial_a\phi\partial^b\phi\eta+\frac{1}{2}(h)],\label{eqn:a68}
\end{eqnarray}here ${}^*T_a$ is the   flat space tensorial component of the one form ${}^*T$.  Now using symmetries of the Einstein's tensor $G^{ab}$, we can break this equation to a symmetric part and an antisymmetric part, and can write
\begin{eqnarray}
G^b{}_a\eta&=&-\kappa[ \frac{i}{8}\{\overline{\Psi}(\gamma^b D_a+\gamma_a D^b)\Psi-(\overline{D_a\Psi}\gamma^b+\nonumber\\&{}&\overline{D^b\Psi}\gamma_a)\Psi\}\eta-\frac{g}{16}\overline{\Psi}\gamma_5(\gamma_a {}^*T^b+\gamma^b {}^*T_a)\Psi\eta\nonumber\\&{}&+f\partial_a\phi\partial^b\phi\eta+\frac{1}{2}(h)\eta\delta^b{}_a],\label{eqn:e1}\\0&=&[\frac{1}{2}  \overline{\nabla}_\nu\overline{\Psi}\{\frac{\sigma^b{}_{a}}{2}, \gamma^\nu\} \Psi+\frac{i}{2}\{\overline{\Psi}(\gamma^b D_a-\gamma_a D^b)\Psi\nonumber\\&{}&-(\overline{D_a\Psi}\gamma^b-\overline{D^b\Psi}\gamma_a)\Psi\}\nonumber\\&{}& -\frac{g}{4}\overline{\Psi}\gamma_5(\gamma_a {}^*T^b-\gamma^b {}^*T_a)\Psi]\eta.\label{eqn:e2}
\end{eqnarray}

Again using (\ref{eqn:abc17}), (\ref{eqn:a58}) and (\ref{eqn:a59}),  equations (\ref{eqn:abc14}) and (\ref{eqn:abc140}) reduce to
\begin{eqnarray}&{}&    d(\mbox{$\mathcal{R}$}-\beta\phi^2-\frac{1}{2}\kappa m_\Psi \overline{\Psi} \Psi)=-\frac{g}{4}\overline{\Psi}\gamma_5\gamma\Psi\nonumber\\\mbox{or,}&{}&\kappa d[\frac{g}{4}{}^*(\overline{\Psi}\gamma_5\gamma\Psi\wedge T)-f{}^*(d\phi\wedge{}^*d\phi)\nonumber\\&{}&+2h-\frac{\beta}{\kappa}\phi^2]=-\frac{g}{4}\overline{\Psi}\gamma_5\gamma\Psi,\label{eqn:abc141}
    \\ &{}&\frac{2}{\kappa}\beta \phi + f^\prime(\phi)d\phi\wedge{}^*d\phi -h^\prime(\phi)
\eta+2fd{}^*d\phi\nonumber\\&{}&\hspace{2mm}=-2\phi[\frac{i}{2}\{\overline{\psi}{}^*\gamma\wedge D\psi+\overline{D\psi}\wedge{}^*\gamma\psi\}\nonumber\\&{}&\hspace{6mm}-\frac{g}{4}\overline{\psi}\gamma_5\gamma\psi\wedge T+m_\psi \overline{\psi}\psi\eta]\nonumber\\&{}&\hspace{2mm}=0.\label{eqn:abc142}
\end{eqnarray}
\end{appendix}
                

\end{document}